\documentclass[prd,twocolumn,superscriptaddress,amsfonts,amssymb,amsmath,showpacs]{revtex4-2}
\usepackage{bm}
\usepackage{amsfonts}
\usepackage{latexsym}
\usepackage[latin1]{inputenc}
\usepackage{graphicx}
\usepackage{amsmath}
\usepackage{palatino}
\usepackage{mathpazo}
\usepackage{textcomp}
\usepackage{float}
\usepackage{booktabs}
\usepackage{dcolumn}
\usepackage{booktabs}
\usepackage{multirow}
\usepackage{hyperref}
\hypersetup{
    colorlinks=true,
    linkcolor=purple,
    filecolor=magenta,      
    citecolor=blue
}
\usepackage{amsmath}
\usepackage{xcolor}
\usepackage{orcidlink}
\usepackage[caption=false]{subfig}
\usepackage{commath}
\makeatletter
\long\def\dddddot#1{%
  {\mathop {#1}\limits ^{\vbox to-1.4\ex@ {\kern -\tw@ \ex@ \hbox {\normalfont .....}\vss }}}%
}
\long\def\multidots#1#2{%
  \count@=0
  {{\mathop {#2}\limits ^{\vbox to-1.4\ex@ {\kern -\tw@ \ex@ \hbox {\normalfont %
  \loop%
  \ifnum#1>\count@%
  .%
  \advance\count@ by1%
  \repeat%
  }\vss }}}}%
}
\makeatother

\begin{document}

\color{black}       

\title{Weyl type $f(Q,T)$ gravity observational constrained cosmological models}

\author{Rahul Bhagat\orcidlink{0009-0001-9783-9317}}
\email{rahulbhagat0994@gmail.com}
\affiliation{Department of Mathematics, Birla Institute of Technology and
Science-Pilani,\\ Hyderabad Campus, Hyderabad-500078, India.}
\author{S.A. Narawade\orcidlink{0000-0002-8739-7412}}
\email{shubhamn2616@gmail.com}
\affiliation{Department of Mathematics, Birla Institute of Technology and
Science-Pilani,\\ Hyderabad Campus, Hyderabad-500078, India.}
\author{B. Mishra\orcidlink{0000-0001-5527-3565}}
\email{bivu@hyderabad.bits-pilani.ac.in}
\affiliation{Department of Mathematics, Birla Institute of Technology and
Science-Pilani,\\ Hyderabad Campus, Hyderabad-500078, India.}

\date{\today}

\begin{abstract}
\textbf{Abstract}: In this paper, we have studied the dynamical aspects of the cosmological model of the Universe in the Weyl type $f(Q,T)$ gravity, which is an extension of symmetric teleparallel gravity. The non-metricity scalar $Q$ has been expressed in standard Weyl form and can be determined by a vector field $w_{\mu}$ and the trace of energy momentum tensor denoted as $T$. The logarithmic form of the Hubble parametrization has been incorporated and the best fit values of the free parameters have been determined using $32~CC$ sample points, $1701~Pantheon^{+}$ and $6~BAO$ data points. The present value of the $H_{0}\approx 70.2\pm 4.6$, $H_{0}\approx 68.69_{-0.59}^{+0.67}$ and $H_{0}\approx 69.26_{-0.53}^{+0.57}$ respectively for $CC~Sample$, $CC + Pantheon^{+}$ and $CC + Pantheon^{+} + BAO$ datasets. With the constrained values of the free parameters, the cosmographic parameters are constrained and the present value of each parameter has been noted. The deceleration parameter for $CC~Sample$, $CC + Pantheon^{+}$ and $CC + Pantheon^{+} + BAO$ datasets provides $-0.5221$, $-0.5477$ and $-0.5691$ respectively at present time. We have considered exponential and non-linear form of the Weyl type function $f(Q,T)$ to assess the dynamical behaviour of the model. The accelerating cosmological models show the quintessence behaviour at present time as we get the present EoS parameter values obtained as $\omega\approx -0.7068$ and $\omega\approx -0.6828$ for $CC~Sample$,  $\omega\approx -0.6991$ and $-0.6949$ for $CC + Pantheon^{+}$ and $\omega\approx -0.7001$ and $-0.7084$ for $CC + Pantheon^{+} + BAO$ respectively.
\end{abstract}

\maketitle

\section{Introduction}
In recent cosmological observations \cite{Riess1998, Perlmutter1999, Aghanim2020}, it has been demonstrated that the Universe is expanding at an accelerated rate. The cosmological observations indicate some exotic form of energy may be responsible for this behavior. This form of energy is known as the dark energy (DE). This kind of behaviour of the Universe is according to the availability of high precision observational evidence such as Large Scale Structure (LSS) \cite{Daniel2008}, Baryonic Acoustic Oscillations (BAO) \cite{Eisenstein2005}, the Wilkinson Microwave Anisotropy Probe experiment (WMAP) \cite{Komatsu2003}, Cosmic Microwave Background Radiation (CMBR) \cite{Hinshaw2013, Komatsu2011}, Baryon Oscillation Spectroscopic Survey (BOSS) \cite{Alam2017}, and the Plank collaboration \cite{Ade2016}, etc. According to the results of these observations, in the mass-energy budget of the Universe, DE occupies roughly $68.3\%$ of the entire Universe, whereas the dark matter and ordinary matter respectively has share of $26.8\%$ and $4.9\%$. General Relativity (GR) has limitations to address this issue, so a new gravity or modification is inevitable. GR can be modified either by modifying the geometry or by modifying the matter part by adding exotic matter. The modification in the geometry has shown some significant results in addressing this current behaviour of the Universe.
  
The modification of GR can be performed by changing the gravitational sector, that leads to the modified theories of gravity. Based on Riemannian geometry, GR is described by a Levi-Civita connection. Here, $R$, the curvature, is considered to be the building block of space-time, with vanishing torsion and non-metricity. One of the geometries that can be used to describe GR, besides Riemannian geometry, is that the geometrical part extending to the equivalent torsion formulation. This is often called the teleparallel equivalence of GR (TEGR) \cite{Einstein1928c,Einstein1928d,Arco2004}. In this class of modified gravity, the Weitzenb$\ddot{o}$ck connection which attributes gravity to torsion instead of curvature  known as the teleparallel gravity and subsequently the  $f(\mathcal{T})$ gravity theory, where $\mathcal{T}$ is the torsion scalar \cite{Capozziello201,Cai2016,Capozziello2011}. In addition to these two equivalent representations, the non-metricity $Q$ of the metric can be used to represent the basic geometric variable describing gravitational interaction. This approach is called as symmetric teleparallel gravity proposed by Nester and Yo \cite{Nester1999}, and further developed into $f(Q)$ gravity \cite{Jimenez2018}, where the non-metricity $Q$, defined as a variation in vector length during parallel transport around a close loop.

The cosmological and astrophysical aspects of symmetric teleparallel gravity or $f(Q)$ gravity are receiving increasing attention in recent times among the modified theories of gravity. In an extension of symmetric teleparallel gravity within the framework of metric-affine formalism, Harko et al. \cite{Harko2018} have given the modified $f(Q)$ gravity. Some aspects of $f(Q)$ gravity are available in the literature, see Ref. \cite{Lazkoz2019, Hohmann2019, Soudi2019, Bajardi2020, Lin2021, Frusciante2021, Anagnostopoulos2021, Narawade2022}. A non-minimal coupling in the gravitational action is present in $f(Q,T)$ gravity. Another extension of $f(Q)$ gravity in which the Lagrangian is replaced by an arbitrary function $f$ of the non-metricity $Q$ and the trace of the energy momentum tensor $T$ proposed by Xu et al. \cite{Xu2019}. Najera and Fajardo developed the linear theory of perturbations in $f(Q, T)$ theories \cite{ Najera2022}. Several studies have shown that $f(Q,T)$ gravity can address some of the key issues of the present behaviour of the Universe \cite{Pati2021a, Agrawal2021, Pati2021b, Zia2021, Najera2021, Godani2021, Iosifidis2021, Pradhan2021, Shiravand2022, Sokoliuk2022}. \\

 The extension of $f(Q,T)$ gravity is the Weyl type $f(Q,T)$ which has been recently introduced by Xu et al. \cite{Xu2020}. For the purpose of describing the early and late stages of cosmic evolution, the Weyl type $f(Q,T)$ gravity can be thought of as an alternative technique. The original form of Weyl gravity that unifies gravity and electromagnetism has been proposed \cite{weyl1918c} and recently, Weyl gravity revived to solve the problem of dark matter and DE, or Inflation \cite{Alvarez2017}. Weyl gravity in the presence of a non-minimal matter-curvature coupling is presented by Gomes and Bertolami \cite{Gomes2019}. In Weyl geometry of the vector field $\omega_\mu$, the general field model equations resulting from action variations allows a complete description of the metric tensor of gravitational phenomena in the form of vector fields $\omega_\mu$. The Weyl-type $f(Q,T)$ gravity has some important applications, for example  when the full connection of Weyl conformal gravity is varied instead of the metric only, the resulting vacuum field equations reduce to vacuum Einstein equation with the choice of local units, if and only if the torsion vanishes\cite{Wheeler2014}. Yang et al. \cite{Yang2021} considered the geodesic deviation equation and the Raychaudhuri equation, and studied the evolution of kinematical quantities associated with deformations in the Weyl-type $f(Q,T)$ gravity. In addition, the Newtonian limit of the theory has been investigated and derived the generalized Poisson equation that contains the correction terms from the Weyl geometry.

Here we have considered the Weyl type $f(Q,T)$ gravity with the non-linear and exponential form of $f(Q,T)$. The Hubble parametrization has been done from a well motivated fundamental $\Lambda$CDM cosmology. The model parameters have been constrained with high-precision cosmological data collected from observation, such as $CC$ sample, $Pantheon^+$ dataset and $BAO$ dataset.\\

The paper has been organised as: In Sec. \ref{Section 2}, a complete discussion of Weyl-type $f (Q,T)$ gravity and the gravitational action with its field equations has been given. In Sec. \ref{section 3}, We have performed observational constraints to parameterize $H(z)$ with $CC$, $Pantheon^+$ and $BAO$ dataset. We have presented the cosmological parameters in Sec. \ref{section 4} and analyzed the non-linear and exponential form of $f(Q,T)$ in the Weyl type $f(Q,T)$ gravity in Sec. \ref{section 5}. The energy conditions of both the models are shown in Sec. \ref{section 6}. Finally,  we have given the conclusion of the results in Sec. \ref{section 7}.

\section{Field Equations of Weyl type $f(Q,T)$ gravity}\label{Section 2}

The gravitational action of Weyl type $f(Q,T)$ gravity \cite{Xu2020},
\begin{eqnarray}\label{eq:1}
S &=& \int dx^4 \sqrt{-g} \big[\kappa^2 f(Q,T) -\frac{1}{4} W_{\mu\nu} W^{\mu\nu} -\frac{1}{2}m^{2} \omega_{\mu} \omega^{\mu} \nonumber \\ 
&+&\lambda (R +\bigtriangledown_\alpha \omega^ \alpha-6 \omega_\alpha \omega^ \alpha) +\mathcal{L}_m \big]~,
\end{eqnarray}
 where $\kappa^2=\frac{1}{16\pi G}$, $\mathcal{L}_m$ is the matter Lagrangian with $m$ be the mass of the particle corresponding to the vector field, $R$ is the Ricci scalar constructed from Riemann tensor and the Levi-Civita connection and $g = det(g_{\mu\nu} )$ be the determinant of the metric tensor. The second term represents the standard kinetic term whereas the third term is the mass term of the vector field. Also $W_{\nu\mu}=\triangledown_\mu \omega_\nu-\triangledown_\nu \omega_\mu$.  
 
 The Levi-Civita connection is compatible with the metric in Riemannian geometry, i.e.,$\triangledown_\alpha g_{\mu\nu}=0$. However, in Weyl geometry it is not the same as it contains semi-metric connection.
\begin{align}\label{eq:2}
\tilde Q_{\alpha\mu\nu}&\equiv\tilde\triangledown_\alpha g_{\mu\nu}=\partial_\alpha g_{\mu\nu} -\tilde \Gamma ^\rho _{\alpha \mu}g_{\rho \nu}-\tilde \Gamma ^\rho_{\alpha\nu}g_{\rho\mu}\nonumber\\& = 2\omega_\alpha g_{\mu\nu},
\end{align}
where, $ \tilde \Gamma ^\lambda_{\mu\nu}=\Gamma^{\lambda}_{\mu\nu}+g_{\mu\nu}w^\lambda-\delta_\mu^\lambda w_\nu - \delta_\nu^\lambda w_\mu$ and $\Gamma ^\lambda_{\mu\nu}$ is the Christoffel symbol constructed with respect to the metric $g_{\mu\nu} $. The scalar non-metricity is,
\begin{equation}\label{eq:3}
Q \equiv -g^{\mu\nu} (L^\alpha_{\beta\mu}L^\beta _{\nu\alpha} - L^\alpha _{\beta\alpha} L^\beta _{\mu \nu}),
\end{equation}
where $L^\lambda _{\mu\nu}$ is defined as,
\begin{equation}\label{eq:4}
L^\lambda _{\mu\nu}=-\frac{1}{2}g^{~\lambda\gamma}(Q_{~\mu\gamma\nu}+Q_{~\nu\gamma\mu}-Q_{~\gamma\mu\nu}),
    \end{equation}
From Eqns. \eqref{eq:2}--\eqref{eq:4}, one can get the relation,
\begin{equation}\label{eq:5}
Q=-6 \omega ^2.
\end{equation}
The Weyl type $f(Q,T)$ gravity field equation can be obtained by applying the variation principle to the metric tensor and Weyl vector in Eqn. \eqref{eq:1} as,
\begin{multline}\label{eq:6}
\frac{1}{2}(T_{\mu\nu}+S_{\mu\nu})-\kappa^2 f_T (T_{\mu\nu}+\Theta_{\mu\nu})=-\frac{\kappa^2}{2}g_{\mu\nu}f\\-6\kappa^2 f_Q \omega_\mu \omega_\nu+\lambda(R_{\mu\nu}-6\omega_\mu \omega_\nu+3g_{\mu\nu}\triangledown_\rho \omega^\rho)\\+3g_{\mu\nu}\omega^\rho\triangledown_\rho\lambda-6\omega_{(\mu}Q_{\nu)}\lambda+g_{\mu\nu}\square \lambda- \triangledown_\mu\triangledown_\nu \lambda,
\end{multline}
where,
\begin{equation}\label{eq:7}
T_{\mu\nu}\equiv -\frac{2}{\sqrt{-g}}\frac{\delta(\sqrt{-g}\mathcal{L}_m)}{\delta g^{\mu\nu}}
\end{equation}
and
\begin{equation}\label{eq:8}
f_T\equiv \frac{\partial f(Q,T)}{\partial T},~~~~~~f_Q\equiv \frac{\partial f(Q,T)}{\partial Q},
\end{equation}
respectively. Further, the quantity $\Theta_{\mu\nu}$ can be defined as
\begin{equation}\label{eq:9}
\Theta_{\mu\nu} \equiv g^{\alpha\beta}\frac{\delta T_{\alpha \beta}}{\delta g_{\mu\nu}}=g_{\mu\nu}\mathcal{L}_m-2T_{\mu\nu}-2g^{\alpha\beta}\frac{\delta^2\mathcal{L}_m}{\delta g^{\mu\nu}\delta g^{\alpha\beta}},
\end{equation}\\
where $S_{\mu\nu}$ is the free Proca field rescaled energy momentum tensor in the field equation,
\begin{multline}\label{eq:10}
S_{\mu\nu}=-\frac{1}{4}g_{\mu\nu}W_{\rho\sigma}W^{\rho\sigma}+W_{\mu\rho}W^\rho_\nu\\-\frac{1}{2}m^2 g_{\mu\nu}\omega_\rho \omega^\rho+m^2\omega_\mu \omega_\nu~.
\end{multline}
We consider the isotropic, homogeneous and spatially flat  Friedmann-Lema$\hat{i}$tre-Robertson-Walker (FLRW) space time as, 
\begin{equation}\label{eq:11}
ds^2=-dt^2+a^2(t)(dx^2+dy^2+dz^2),
\end{equation}
$a(t)$ represents the expansion rate along the spatial coordinates. Because of spatial symmetry, the vector field can be assumed in the following form,
\begin{equation}\label{eq:12}
\omega_\nu=[\psi(t),~0,~0,~0]
\end{equation}
So, we get from above $\omega^2=\omega_\nu\omega^\nu=-\psi^2(t)$, with $Q=-6\omega^2=6\psi^2(t)$. In addition, we use a comoving coordinate system that $u^\mu=(-1,~0,~0,~0)$. So, $u^\mu\triangledown_\mu=\frac{d}{dt}$ and $H=\frac{\dot a}{a}$. We consider the Lagrangian of the perfect fluid, $\mathcal{L}_m=p$. As a result, $T_\nu^\mu=diag(-\rho,~p,~p,~p)$ and 
\begin{equation}\label{eq:13}
\Theta^\mu_\nu=\delta^\mu_\nu p-2T^\mu_\nu=diag(2\rho+p,-p,-p,-p)
\end{equation}
For the cosmological case the flat space constraint, and the generalized Proca equation can be represented as
\begin{eqnarray}\label{eq:13.1}
    \dot{\psi} &=& \dot{H} + 2H^{2} + \psi^{2} -3H\psi~,\nonumber\\
    \dot{\lambda} &=& \left(-\frac{1}{6}m^{2} -2\kappa^{2}f_{Q} -2\lambda \right)\psi ~,\nonumber\\
    \partial_{i}\lambda &=& 0~.
\end{eqnarray}

From Eqn. \eqref{eq:6}, the field equations of Weyl type $f(Q,T)$ gravity can be obtained as,
\begin{multline}\label{eq:14}
\kappa^2f_T(\rho+p)+\frac{1}{2}\rho=\frac{\kappa^2}{2}f-\left(6\kappa^2f_Q+\frac{1}{4}m^2\right)\psi^2 \\-3\lambda(\psi^2-H^2)-3\dot\lambda(\psi-H)
\end{multline}
\begin{multline}\label{eq:15}
-\frac{1}{2}p=\frac{\kappa^2}{2}f+\frac{m^2\psi^2}{4}+\lambda(3\psi^2+3H^2+2\dot H)\\ +(3\psi+2H)\dot\lambda+\ddot\lambda   
\end{multline}

\section{Observational constraints}\label{section 3}
\subsection{$H(z)$ Parameterizations}
We use the the log model parameterised form of $H(z)$ \cite{Lemos2018} as,
\begin{equation}\label{eq:16}
 H^2(z)=H^2_{0}\left[A(1+z)^3+B+Cz+D\log (1+z)\right],
 \end{equation}
with $H_0$, $A$, $B$, $C$ and $D$ as free parameters. The normalising factor $H_{0}$ ensures that the free parameters $A$ to $D$ are dimensionless and of order unity. The present observed value of Hubble parameter, $H_{0}=72.1\pm 2.0 kms^{-1}Mpc^{-1}$ \cite{Soltis2020}, $H_{0}=69.8\pm 0.6 kms^{-1}Mpc^{-1}$ \cite{Freedman2021} and $H_{0}=68.3\pm 1.5 kms^{-1}Mpc^{-1}$ \cite{Balkenhol2022}. We use the fundamental $\Lambda$CDM cosmology,
\begin{equation}\label{eq:17}
H(z)=H_0\left[\Omega_m(1+z)^3+(1-\Omega_m)\right]^{\frac{1}{2}},
\end{equation}
where $\Omega_{m}$ represents the current total matter density. Eqn. \eqref{eq:16} is applicable at low redshifts, where the contribution of photon and neutrino to the energy density can be ignored. The parametrization of Eqn. \eqref{eq:16} reproduces if,
\begin{eqnarray}
A &=& \Omega_m~,~~~~B=(1-\Omega_m),\nonumber\\
C &=& D = 0.\label{eq:18}
\end{eqnarray}
Using Eqn. \eqref{eq:18}, Eqn. \eqref{eq:16} can reduce to $\Lambda$CDM cosmology. According to $\Lambda$CDM model, the value of equation of state (EoS) parameter,  $\omega = \frac{p}{\rho}=-1$ in the DE phase. The EoS parameter has been in frequent use to develop the DE model and one can express as,
\begin{equation}\label{eq:19}
\omega (z)=\omega_0+\omega_a\left(\frac{z}{1+z}\right),
\end{equation}
where $\omega_0$ and $\omega_a$ are constants. Assuming $\omega_a=0$ and $\omega_0=-1$, the EoS parameter reduces to $\Lambda$CDM. With this Eqn. \eqref{eq:16} can be rewritten as,
 \begin{multline}\label{eq:20}
 H^2(z)=H_0^2\left[\Omega_m(1+z)^3+\Omega_k(1+z)^2\right.\\
 \left.+\Omega_{DE}(1+z)^{-3(1+\omega_0+\omega_a)e^{-3\omega_a\frac{z}{1+z}}}\right],
\end{multline}
where $\Omega_k$ is an arbitrary curvature and  $\Omega_{DE}=1-\Omega_m-\Omega_k$.

\subsection{$Hubble$ Dataset}
 We use the $32~Hubble$ measurements from the range $0.07 \leq z \leq 1.965$ given in Table-\ref{table:III} obtained by the differential age technique, which is independent of the fiducial model. Also known as cosmic chronometers $(CC)$ \cite{Jimenez2002}, we calculate the mean values of the model parameters $H_{0}$, $A$, $B$, $C$, and $D$ by minimising the chi-square value. The $Hubble~(CC)$ chi-square ($\chi^2$) can be calculated with,
 \begin{equation}\label{eq:21}
 \chi^{2}_{CC}(P_s)=\sum_{i=1}^{32} \frac{[H_{th}(z_i,P_s)-H_{obs}(z_i)]^2}{\sigma_H^2(z_i)},
 \end{equation}
where $H_{obs}(z_i)$ represents the observed Hubble parameter values, $H_{th}(z_i,P_s)$ represents the Hubble parameter with the model parameters and $\sigma_H^2(z_i)$ is the standard deviation.

\subsection{$Pantheon^{+}$ Dataset}
The $Pantheon^{+}$ dataset consists of $1701$ light curves of $1550$ distinct Type Ia supernovae (SNe Ia), in the redshift range  $0.00122\leq z \leq 2.2613$  \cite{Brout2022}. According to the measurements of type Ia supernovae, the accelerated expansion of the Universe has been reported. As a result of this expansion, when the  stellar objects emit light, they are redshifted. The luminance of a star object determines the luminosity distance. The distance modulus is used to calculate the distance from the supernova. By contrasting the observed and hypothesised values of the distance moduli, the model parameters are to be fitted. The $\chi^2_{Pantheon^{+}}$ has been given as,
 \begin{equation}\label{eq:22}
 \chi^2_{Pantheon^{+}}(z,\theta)=\sum_{i=1}^{1701}\frac{[\mu(z_i,\theta)_{th}-\mu(z_i)_{obs}]^2}{\sigma^2_\mu(z_i)},
 \end{equation}
 where $\sigma^2_{\mu}(z_i)$ is the standard error in the observed value and $\mu_{th}$ is the theoretical distance modulus. It has been define as,
 \begin{equation}\label{eq:23}
 \mu_{th}=\mu(D_L)=m-M=5\log_{10}D_L(z)+\mu_0,
 \end{equation}
$\mu_0$ is the nuisance parameter and $D_L(z)$ is the dimensionless luminosity distance defined as,
\begin{equation}\label{eq:24}
D_L(z)=(1+z)\int_{0}^{z}\frac{1}{E(z^*)}dz^*.
\end{equation}
the dimensionless parameter, $E(z)=\frac{H(z)}{H_0}$ and $z^*$ is the change of variable that has been defined from  $0$ to $z$.\\
\subsection{$BAO$ Dataset}
The results from SDSS, 6dFGS, and Wiggle Z surveys at different redshifts are available for $BAO$ data. To obtain $BAO$ constraints, the following expressions can be used for the measurable quantities,
\begin{eqnarray}
d_{A}(z_{*}) = \int_{0}^{z_{*}}\frac{d\tilde{z}}{H(\tilde{z})}, \label{eq:25}\\
D_{V}(z) = \left[\frac{(d_{A}(z))^{2}z}{H(z)}\right]^{\frac{1}{3}}. \label{eq:26}
\end{eqnarray}
The distance between comoving angular diameters is $d_{A}(z)$, and the dilation scale is $D_{V}(z)$. The chi-square for $BAO$ datasets can be given as,
\begin{equation}\label{eq:27}
\chi _{BAO}^{2}=X^{T}C^{-1}X.
\end{equation}
with $X$ depends on the survey considered and the inverse of covariance matrix $C$ is as in Refs.\cite{Percival2010, Blake2011, Giostri2012}.

\subsection{$MCMC$ analysis}
In order to explore the parameter space, we employ the Python emcee program \cite{Foreman-Mackey2013} and the $MCMC$ sampling technique. Upon minimization of a total $\chi^2$, it will produce the best fits of $H_{0}$, $A$, $B$, $C$ and $D$. It is note that, while estimating the parameters, the normalising constant will not be calculated. With the use of prior and likelihood estimates, the posterior parameter distributions will be calculated. We have used here the observational $CC$ samples, that contains $32$ data points, $Pantheon^+$ samples, that have 1701 data points and $6~-$ BAO distance data points. Further, we have considered separately $CC$, $CC+pantheon^+$, and $CC + pantheon^++ BAO$ to constrain the parameters of the Weyl type $f(Q,T)$ model. 

In FIG.--\ref{fig:I}, we have presented the error bar plots for both $CC$ sample (Upper Panel) and the distance modulo curve (Lower Panel). The solid lines red, blue and pink respectively denotes for the $CC$ sample, $CC+Pantheon^{+}$ sample and $CC+Pantheon^{+}+BAO$ sample. The broken black line denotes for the $\Lambda$CDM model. One can observe that all these lines are traversing in the middle of the error bars.  In FIG.--\ref{fig:II}, the contour plots for $1-\sigma$ and  $2-\sigma$ confidence levels are shown using the $CC$, $Pantheon^{+}$ and $BAO$ datasets. The values obtained for the parameters are given in TABLE--\ref{table:I}.
\begin{widetext}

\begin{table}[H]
\renewcommand\arraystretch{1.5}
\centering 
\begin{tabular}{|c|c|c|c|} 
\hline\hline 
~~~Coefficients~~~& \textit{CC} Sample  & ~~~\textit{CC} + \textit{Pantheon$^+$}~~~ & ~~~$CC + Pantheon^+ + BAO$~~~\\ [0.5ex] 
\hline\hline
$H_{0}$ & 70.2 $\pm$ 4.6 &  $68.69_{-0.59}^{+0.67}$ & $69.26_{-0.53}^{+0.57}$ \\
\hline
$A$ & 0.297 $\pm$ 0.04 &  $0.285^{+0.050}_{-0.048}$ & $0.264^{+0.039}_{-0.036}$ \\
\hline
$B$ & 0.66$^{+0.11}_{-0.13}$ &  $0.689^{+0.071}_{-0.067}$ & $0.698^{+0.070}_{-0.071}$ \\
\hline
$C$ & 0.02 $\pm$ 0.011 &  $0.012^{+0.98}_{-1.1}$ & $0.012 \pm 0.71$ \\[0.5ex] 
\hline 
$D$ & ~~0.0037 $\pm$ 0.0019~~ &  $0.014^{+1.1}_{-0.99}$ & $0.0025^{+0.81}_{-0.61}$ \\[0.5ex] 
\hline 
\end{tabular}
\caption{Marginalized constraints of the parameters using $CC$, $Pantheon^+$, and $BAO$ datasets.} 
\label{table:I} 
\end{table}
\begin{figure}[H]
\centering
\includegraphics[scale=0.5]{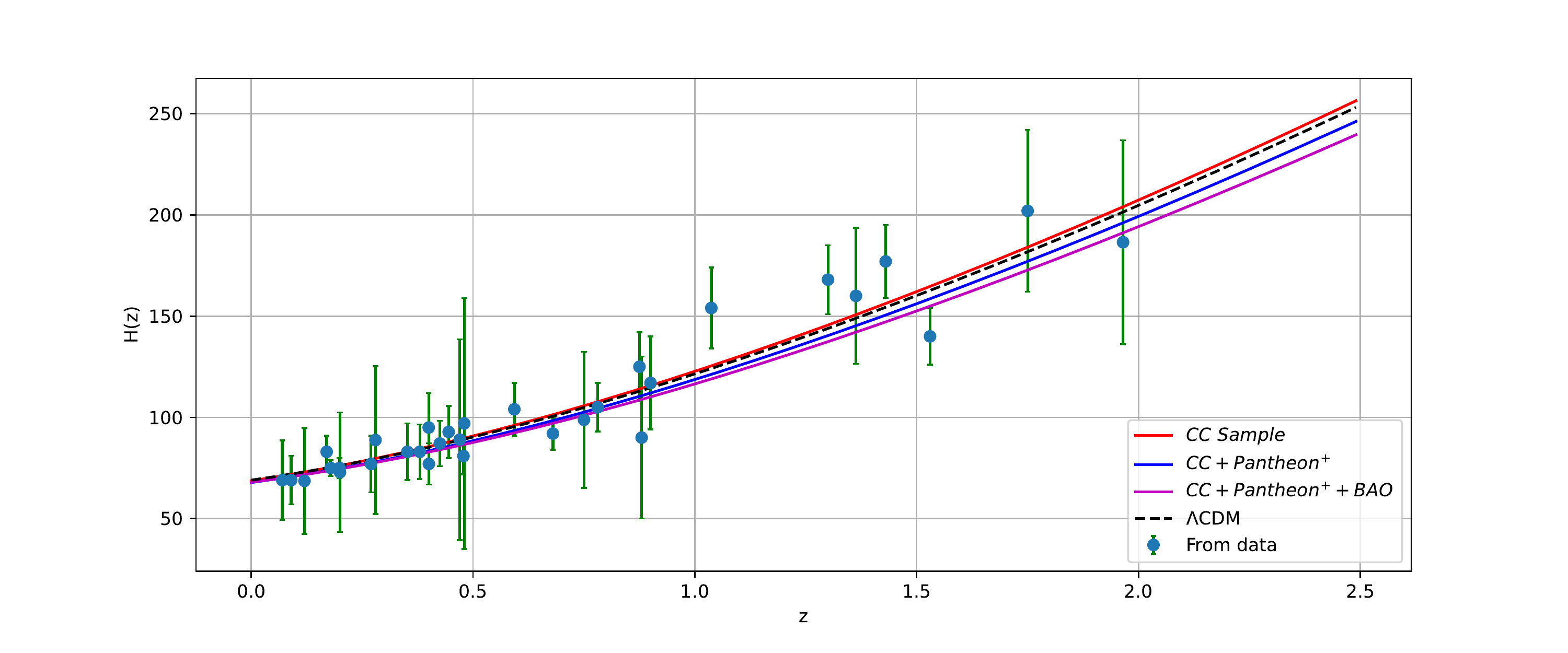}
\includegraphics[scale=0.5]{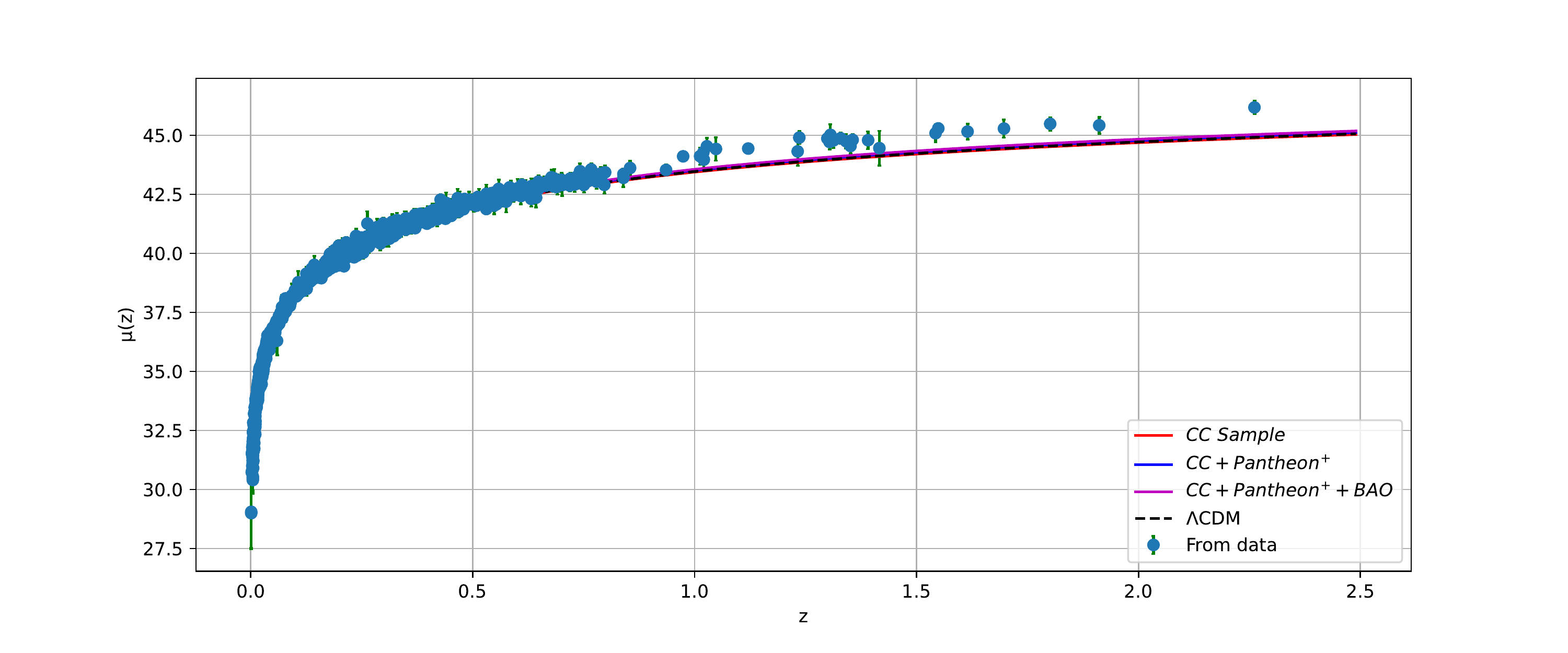}
\caption{Error bar for $CC$ sample (Upper Panel), $Pantheon^{+}$ dataset (Lower Panel) with best-fit values obtained in TABLE-\ref{table:I}}
\label{fig:I}
\end{figure}
\begin{figure}[H]
\centering
\includegraphics[width=12cm,height=12cm]{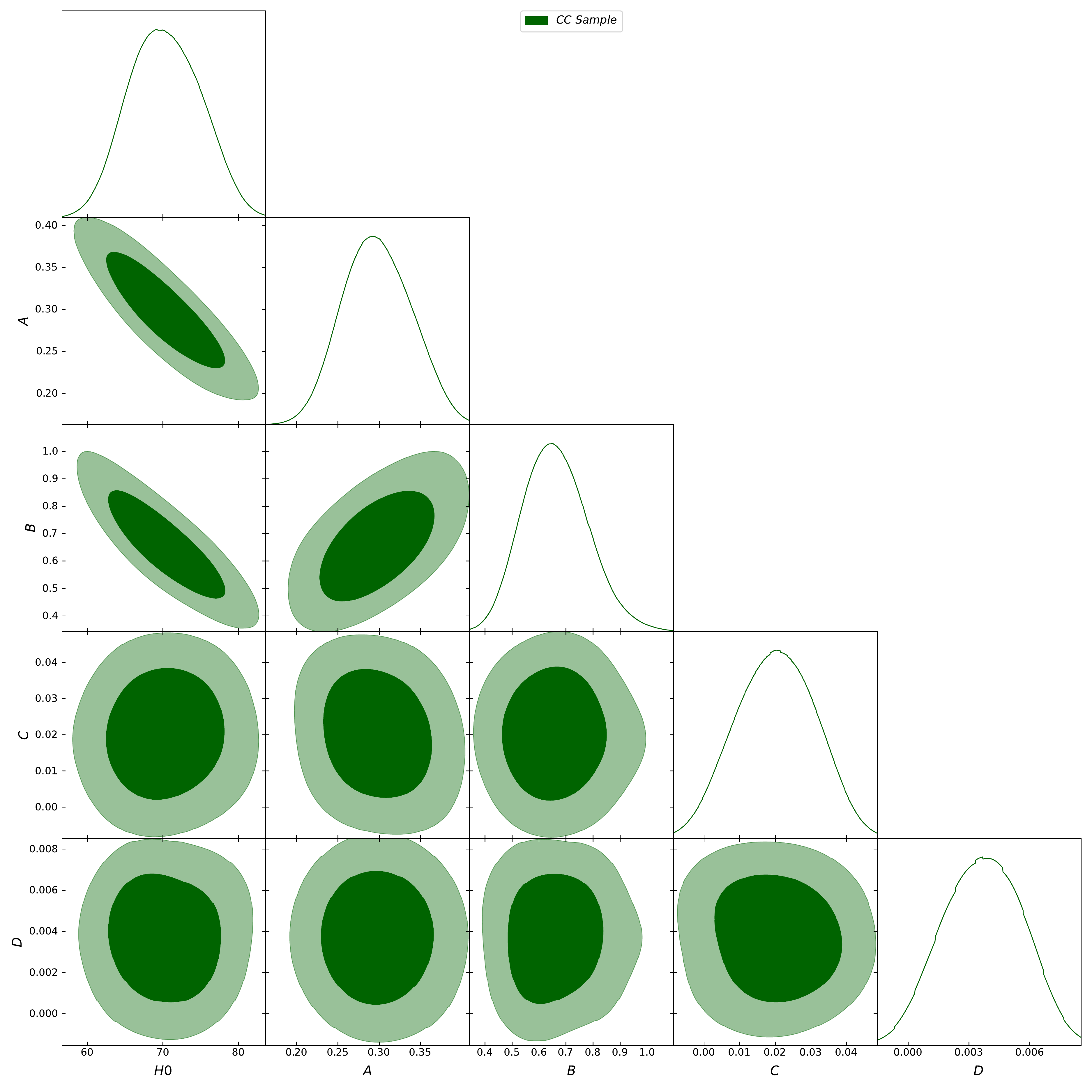}
\includegraphics[width=12cm,height=12cm]{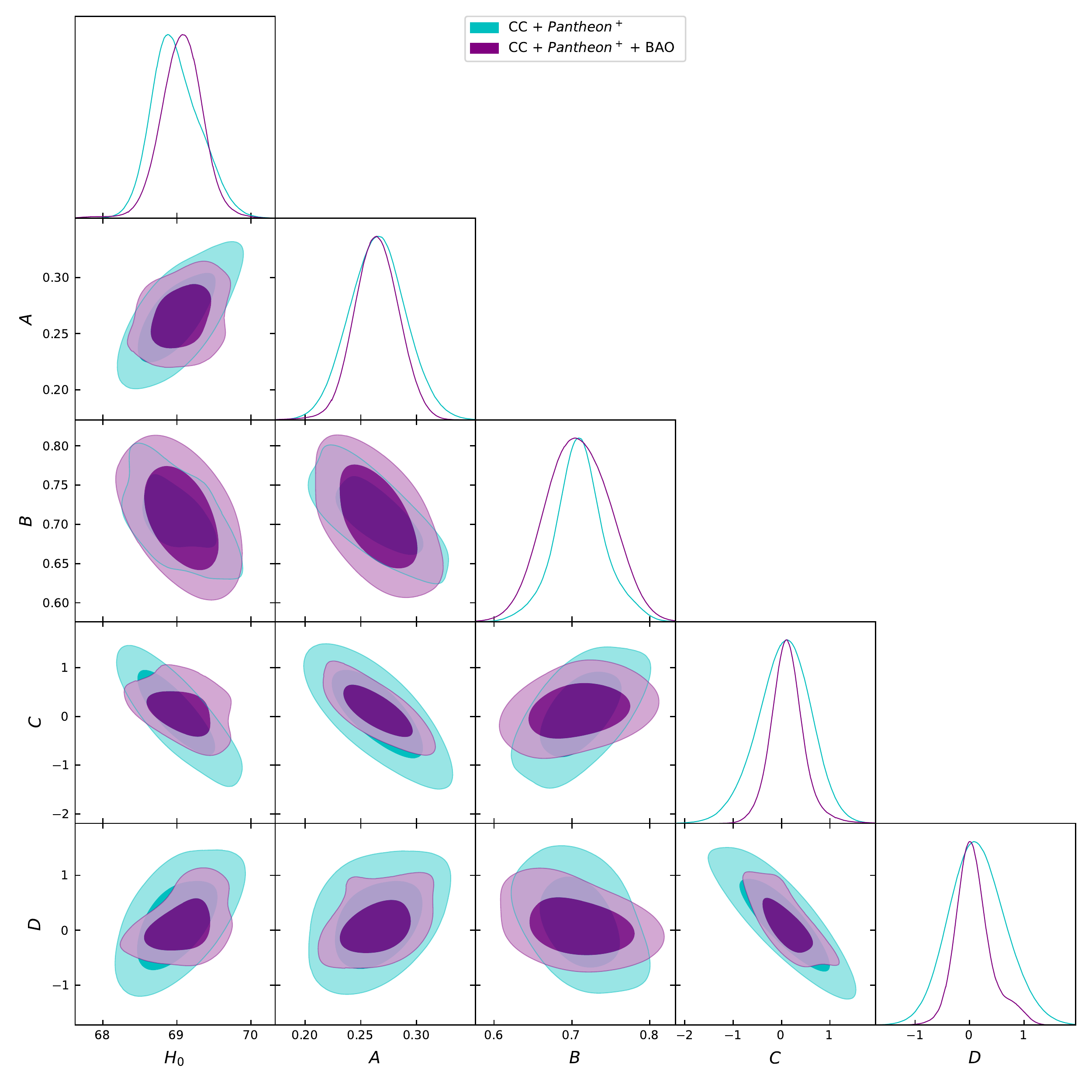}
\caption{The contour plots with $1-\sigma$ and $2-\sigma$ errors for the parameters $H_{0}$, $A$, $B$, $C$ and $D$. Additionally, it contains the parameter values that better match the $32-$ points of $CC$ sample, $1701-$ light curves from $Pantheon^{+}$ dataset and $6-~BAO$ distance dataset.}
\label{fig:II}
\end{figure}
\end{widetext}

\section{COSMOGRAPHIC PARAMETERS}\label{section 4}
\begin{figure}[H]
\centering
\includegraphics[scale=0.5]{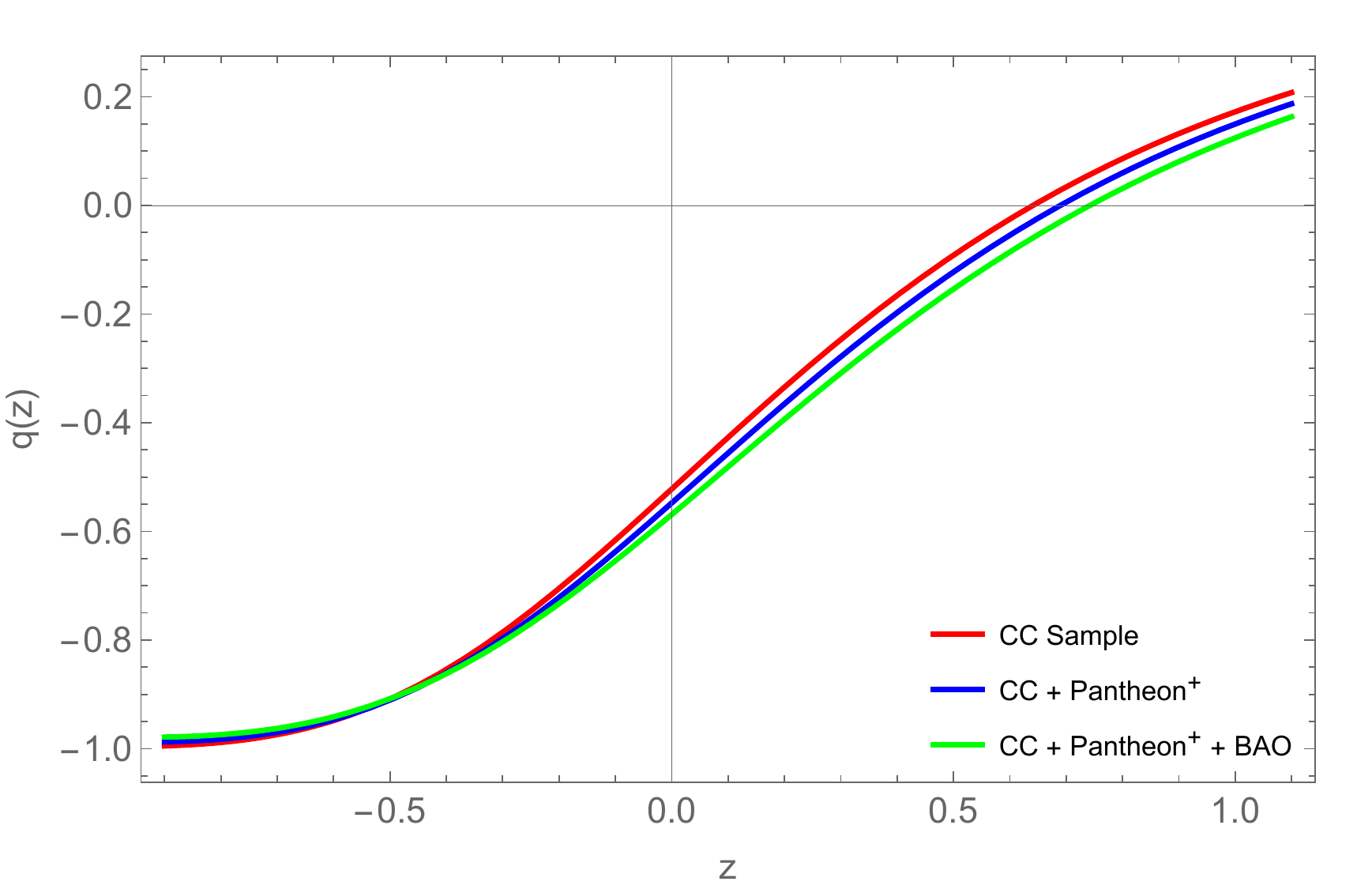}
\caption{Behaviour of deceleration parameter in redshift.}
\label{fig:III}
\end{figure}

We shall analyse the behaviour of the cosmographic parameters which are required to frame the cosmological models. From several cosmological datasets, we have constrained the present value of the Hubble parameter as described in TABLE--\ref{table:I}. The Hubble function is monotonically decreasing with the decrease in redshift value (FIG.--\ref{fig:I}). The deceleration parameter ($q$), the jerk parameter ($j$) and the snap parameter ($s$) can be derived from the Hubble parameter. The deceleration parameter describes the accelerating or decelerating behaviour of the model and can be obtained as, $q=-1-\frac{\dot H}{H^2}$. Considering the constrained values of the Hubble parameter from the datsets, we have plotted the deceleration parameter as in FIG.--\ref{fig:III}. It has been observed that in all three cases, the value of $q$ are decreasing with the decrease in redshift and show the transient behaviour i.e. $q>0$ at early time to $q<0$ at late time. However at sufficient late times it approaches to $-1$ justifying the $\Lambda$CDM behaviour. The present value $q_0$ and the redshift value ($z_t$) at which the transition occurred are given in TABLE--\ref{table:II}. Some of the recent measurements as in Ref. \cite{Gruber2014},  $q_{0} = -0.528_{-0.088}^{+0.092}$  and the transition from deceleration to acceleration occurs at $z_{t} = 0.60_{-0.12}^{+0.21}$ \cite{Yang2020,Capozziello2015} . The present value of deceleration parameter obtained are also compatible as in Ref. \cite{Hernandez-Almada2019,Basilakos2012} and the value at which $q$ changes from positive and negative $(z_t)$ are in the range as in Refs. \cite{Roman-Garza2019,Jesus2020}. 

The state finder diagnostic pair $(j,s)$ describes the geometrical behaviour of the DE cosmological models, specifically it distinguishes various DE models \cite{Sahni2003}. Both the jerk and snap parameter can be obtained with higher order derivative of the scale factor as,
\begin{table}[H]
\renewcommand\arraystretch{2}
\centering
\begin{tabular}{|c|c|c|c|c|} 
\hline \hline
~~Dataset/Parameters~~ & ~~$q_{0}$~~ & ~~$j_{0}$~~ & ~~$s_{0}$~~ & ~~$z_{t}$~~\\
\hline\hline
~~$CC~Sample$~~ & ~~$-0.5221$~~ & ~~$0.9733$~~ & ~~$0.0087$~~ & ~~$0.6411$~~\\ 
~~$CC + Pantheon^{+}$~~ & $-0.5477$ & $0.9661$ & $0.0108$ & $0.6898$ \\
~~$CC + Pantheon^{+} + BAO$~~ & $-0.5691$ & $0.9485$ & $0.160$ & $0.7421$\\
\hline
\end{tabular}
\caption{Present values of cosmographic parameters and the transition point using for best-fit values obtained in TABLE--\ref{table:I}}
\label{table:II}
\end{table}
\begin{eqnarray}
j &=& \frac{\dddot a}{aH^3}~,\nonumber\\
s &=& \frac{j-1}{3(q-\frac{1}{2})}~,\label{eq:28}
\end{eqnarray}
\begin{figure}
\centering
\includegraphics[scale=0.5]{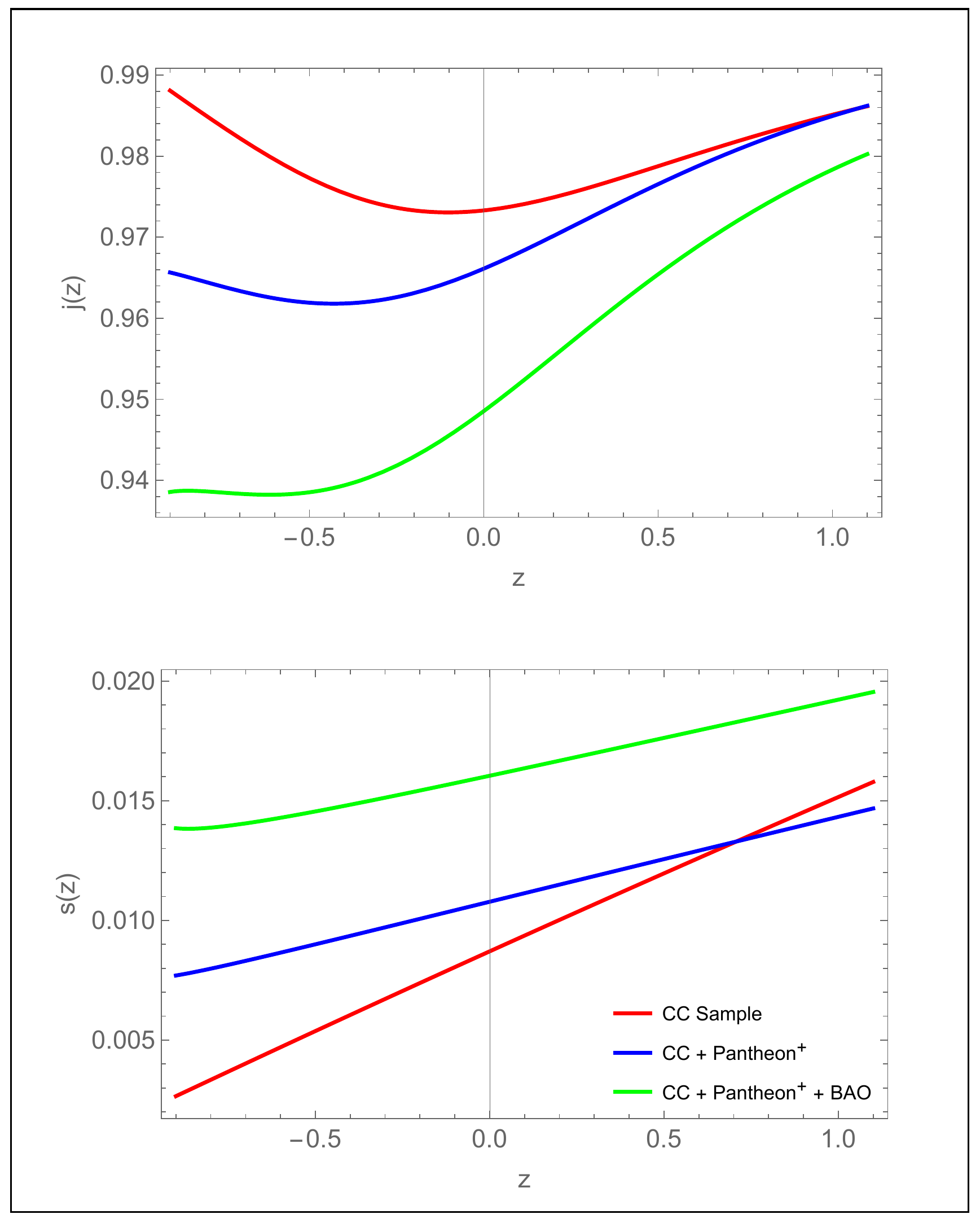}
\caption{Behaviour of jerk parameter(Upper Panel) and snap parameter (Lower Panel) in redshift for the best-fit values mentioned in TABLE-\ref{table:I}.}
\label{fig:IV}
\end{figure}
Sahni et al. \cite{Sahni2003,Alam2003} provides the following for the values of the state finder pair as: (i) $(j=1,~s=0)\rightarrow \Lambda CDM$; (ii) $(j<1,~s>0)$ $\rightarrow $ Quintessence; (iii) $(j>1,~s<0) \rightarrow$ Chaplygin Gas; and  (iv) $(j=1,~s=1) \rightarrow$ SCDM. FIG.--\ref{fig:IV} illustrates the evolution of $j$ and $s$ in redshift using $CC$, $Pantheon^{+}$ and $BAO$ datasets. The present value of jerk parameter, snap parameter and the transition value is given in TABLE-\ref{table:II}. Since the analysis is providing the diagnostic pair as $(j<1,~s>0)$, we can realise the quintessence behavior of the Universe at present time for the constrained values. After obtaining the behaviour of the cosmographic parameters, we shall now frame the cosmological models of the Universe in the Weyl type $f(Q,T)$ gravity with some functional form of $f(Q,T)$ in the following section.

\section{The Models}\label{section 5}
 In order to study the dynamical behaviour, the pressure and energy density as expressed in Eqn. \eqref{eq:14} and Eqn. \eqref{eq:15} are to be explicitly describe in cosmic time. To do that some functional form of $f(Q,T)$ is required. Therefore, we consider two well-motivated functional form of $f(Q,T)$ as: (i) Exponential form and (ii) Non-linear form as below. In addition to know the evolutionary behaviour of the Universe specifically in the DE phase, the present value of EoS parameter, which is the ratio of pressure and energy density is required. Some of the recent observational constrained present value of EoS parameter are: Supernovae Cosmology Project, $\omega = -1.035_{-0.059}^{+0.055}$ \cite{Amanullah2010}; WAMP+CMB, $\omega = -1.079_{-0.089}^{+0.090}$ \cite{Hinshaw2013}; Plank 2018, $\omega = -1.03\pm 0.03$ \cite{Aghanim2020}, $\omega = -1.29_{-0.12}^{+0.15}$ \cite{Valentino2016}, $\omega = -1.3$ \cite{Vagnozzi2020} and $\omega = -1.33_{-0.42}^{+0.31}$ \cite{Valentino2021}.\\

\noindent\textbf{Model--I}: The exponential form given as, $f(Q,T)= \alpha_{1}H_{0}^{2}e^{\frac{\mu Q}{6H_{0}^{2}}}+\frac{\beta_{1}}{6\kappa^{2}}T$, where $\alpha_{1}$, $\beta_{1}$ and $\mu$ are model parameters. The model leads to a complex cosmological dynamics, involving larger deviations from the standard $\Lambda$CDM model.  To note, in the field equations of Weyl type $f(Q,T)$ gravity, $\psi(t)$ is the wave function related to Hubble parameter as given in Eqn. \eqref{eq:13.1} and it provides one of the value of  $\psi$ as $H$ \cite{Xu2020}. Substituting the exponential form of $f(Q,T)$, Eqn. \eqref{eq:14} and Eqn. \eqref{eq:15} can be reduced and also the EoS parameter can be obtained using $\omega=\frac{p}{\rho}$ as, 
\begin{widetext}

\begin{eqnarray}
p &= -\frac{3\left(12\beta_{1}\lambda\dot{H}+24\lambda\dot{H}+2 \alpha_{1}\beta_{1}\kappa^{2}\mu H^{2}e^{\frac{\mu H^{2}}{H_{0}^{2}}}+2\alpha_{1}\beta_{1}H_{0}^{2}\kappa^{2}e^{\frac{\mu H^{2}}{H_{0}^{2}}}+6\alpha_{1}H_{0}^{2}\kappa^{2}e^{\frac{\mu H^{2}}{H_{0}^{2}}}+2\beta_{1}m^{2}H^{2}+3m^{2} H^{2}+36\beta_{1}\lambda H^{2}+72\lambda  H^{2}\right)}{2 \left(2\beta_{1}^2+9\beta_{1}+9\right)}~,\label{eq:29}\\
\rho &= -\frac{3\left(4\beta_{1}\lambda\dot{H}+6\alpha_{1}\beta_{1}\kappa^{2}\mu H^{2}e^{\frac{\mu H^{2}}{H_{0}^{2}}}-2\alpha_{1}\beta_{1}H_{0}^{2}\kappa^{2}e^{\frac{\mu H^{2}}{H_{0}^{2}}}+12\alpha_{1}\kappa^{2}\mu H^{2} e^{\frac{\mu H^{2}}{H_{0}^{2}}}-6\alpha_{1}H_{0}^{2}\kappa^{2} e^{\frac{\mu H^2}{H_{0}^{2}}}+2\beta_{1}m^{2}H^{2}+3m^{2} H^{2}+12\beta_{1}\lambda H^{2}\right)}{2\left(2\beta_{1}^2+9 \beta_{1}+9\right)}~, \label{eq:30}
\end{eqnarray}
\begin{eqnarray}
\omega &=& -\frac{12(\beta_{1}+2)\lambda\dot{H}+H^{2}\left(2\alpha_{1}\beta_{1}\kappa^{2}\mu e^{\frac{\mu H^{2}}{H_{0}^{2}}}+36(\beta_{1}+2) \lambda+(2\beta_{1}+3)m^{2}\right)+2\alpha_{1}(\beta_{1}+3)H_{0}^{2} \kappa^{2}e^{\frac{\mu H^{2}}{H_{0}^{2}}}}{-4\beta_{1}\lambda\dot{H}+H^{2} \left(-6\alpha_{1}(\beta_{1}+2) \kappa^{2}\mu e^{\frac{\mu H^{2}}{H_{0}^{2}}}-12\beta_{1}\lambda+(2\beta_{1}+3)\left(-m^{2}\right)\right)+2\alpha_{1}(\beta_{1}+3)H_{0}^{2}\kappa^{2} e^{\frac{\mu H^{2}}{H_{0}^{2}}}}~.
\label{eq:31}
\end{eqnarray}
\end{widetext}

\begin{figure}[H]
\centering
\includegraphics[scale=0.5]{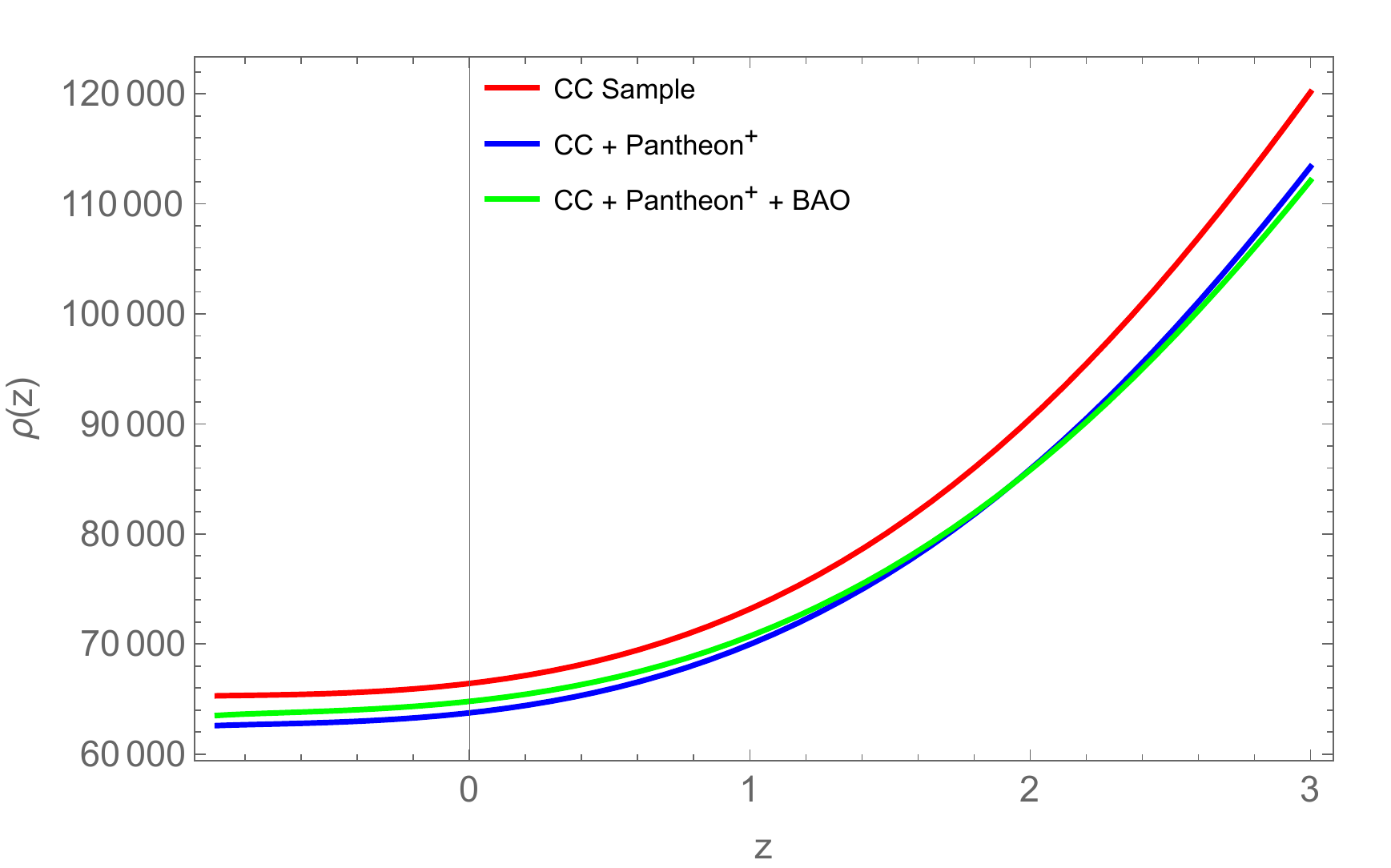}
\caption{Evolution of energy density in redshift. The parameter scheme: $\alpha_{1}=-70$, $\beta_{1}=-10$, $\mu=0.01$, $m=0.98$, $\lambda=\kappa^{2}=1$.}
\label{fig:V}
\end{figure}
\begin{figure}[H]
\centering
\includegraphics[scale=0.5]{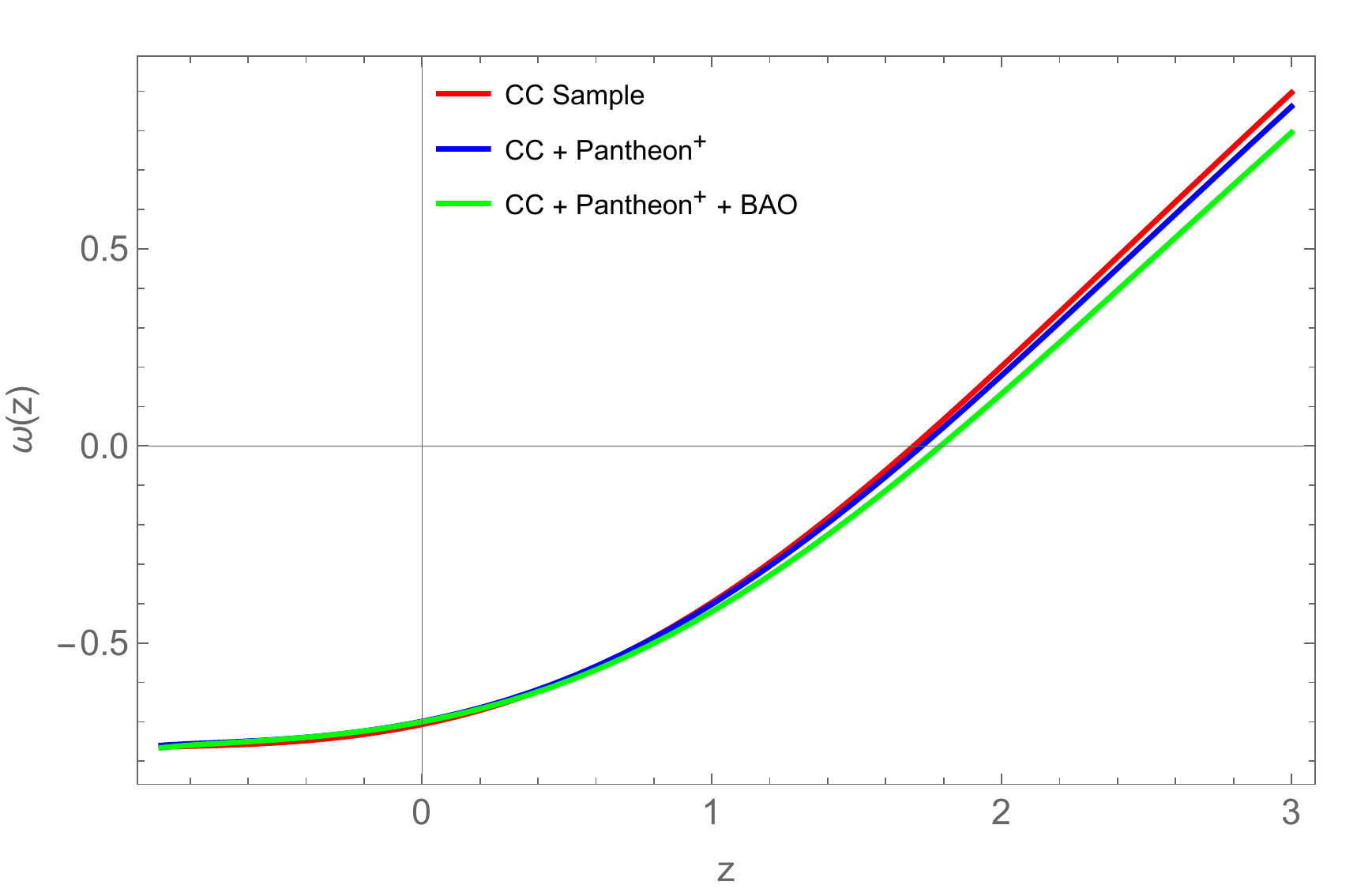}
\caption{Evolution of EoS parameter in redshift. The parameter scheme: $\alpha_{1}=-70$, $\beta_{1}=-10$, $\mu=0.01$, $m=0.98$, $\lambda=\kappa^{2}=1$.}
\label{fig:VI}
\end{figure} 

\begin{figure}[H]
\centering
\includegraphics[scale=0.5]{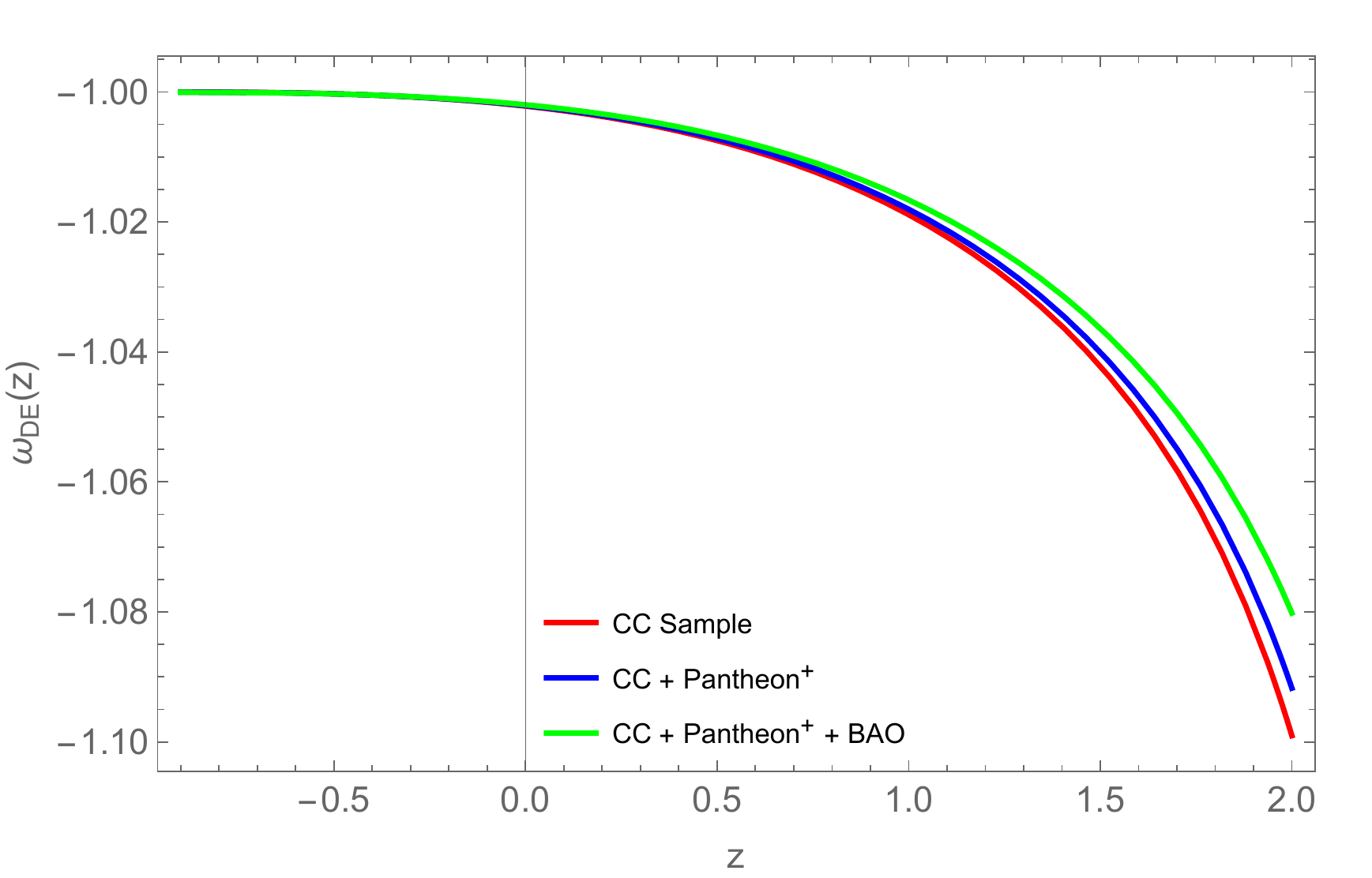}
\caption{Evolution of EoS parameter for dark energy in redshift. The parameter scheme: $\alpha_{1}=-70$, $\beta_{1}=-10$, $\mu=0.01$, $m=0.98$, $\lambda=\kappa^{2}=1$.}
\label{fig:DEI}
\end{figure}

The parameters $\alpha_{1}$, $\beta_{1}$ and $\mu$ determine the evolutionary behavior of energy density and EoS parameter. We have used the constrained values of the Hubble parameter and free parameters from different datasets as given in TABLE--\ref{table:I} and considered the value of model parameter in such a manner that the energy density remains positive throughout the evolution [FIG.--\ref{fig:V}].  The energy density shows decreasing behavior from early epoch to later epoch. The EoS parameter shows the transient behaviour with positive $\omega$ at early time to negative value at late time. For the constrained values obtained from the datasets, the present value of $\omega$ are obtained respectively as, $\omega\approx -0.7068$, $\omega\approx -0.6991$ and $\omega\approx -0.7001$ for $CC~Sample$, $CC + Pantheon^{+}$ and $CC + Pantheon^{+} + BAO$ datasets [FIG.--\ref{fig:VI}].  The behaviour of EoS parameter for dark energy is shown in FIG. --\ref{fig:DEI}. The behaviour is increasing and remaining negative throughout the evolution. The model shows the quintessence behaviour at present time.

\noindent\textbf{Model--II} The non-linear form given as, $f(Q,T)=-Q+\frac{\alpha_{2}}{6H_0^2\kappa^2}QT$, where $\alpha_{2}$ is a model parameter and $H_0$ is the value of Hubble parameter at $z=0$. The non-linear model leads to a good description of the standard cosmological model at small redshift. It allows an appropriate choice of the model parameters in the construction of a large number of accelerating scenarios that inlcudes the de-Sitter type expansions. From the Eqn. \eqref{eq:14} and \eqref{eq:15} using non-linear model, we obtain the expression of $p$, $\rho$ and $\omega$ of the Universe as,

\begin{widetext}

\begin{eqnarray}
 p &=& -\frac{H_{0}^{2}\left(-24\alpha_{2}H^{4}\kappa^2+24\alpha_{2}H^{4}\lambda +2\alpha_{2}H^{4}m^{2}-12H^{2}H_{0}^{2}\kappa^2+24H^{2}H_{0}^{2}\lambda+H^2H_{0}^{2}m^2+8H_{0}^{2}\lambda\dot{H}+8\alpha_{2}H^{2}\lambda\dot{H}\right)}{2\left(8\alpha_{2}^{2}H^{4}+4\alpha_{2}H^{2}H_{0}^{2}+H_{0}^{4}\right)}~,\nonumber\\
 \label{eq:32}\\
 \rho &=& -\frac{H_{0}^{2}\left(24\alpha_{2}H^{4}\kappa^2-120\alpha_{2}H^{4}\lambda -2\alpha_{2}H^{4}m^2-12H^{2}H_{0}^{2}\kappa^2+H^{2}H_{0}^{2}m^{2}-40 \alpha_{2}H^2\lambda\dot{H}\right)}{2 \left(8\alpha_{2}^{2}H^{4}+4\alpha_{2}H^2 H_{0}^{2}+H_{0}^{4}\right)}~,\label{eq:33}\\
 \omega &=& -\frac{H^2\left(-24\lambda \left(\alpha_{2}H^2+H_{0}^2\right)-\left(\left(m^2-12\kappa^2\right)\left(2\alpha_{2}H^2+H_{0}^2\right)\right)\right)-8\lambda\dot{H}\left(\alpha_{2}H^2+H_{0}^2\right)}{H^2\left(120\alpha_{2}H^2\lambda -\left(m^2-12 \kappa ^2\right)\left(H_{0}^2-2 \alpha_{2}H^2\right)+40\alpha_{2}\lambda\dot{H}\right)}~.\label{eq:34}
\end{eqnarray}
\end{widetext}

 \begin{figure}
\centering
\includegraphics[scale=0.5]{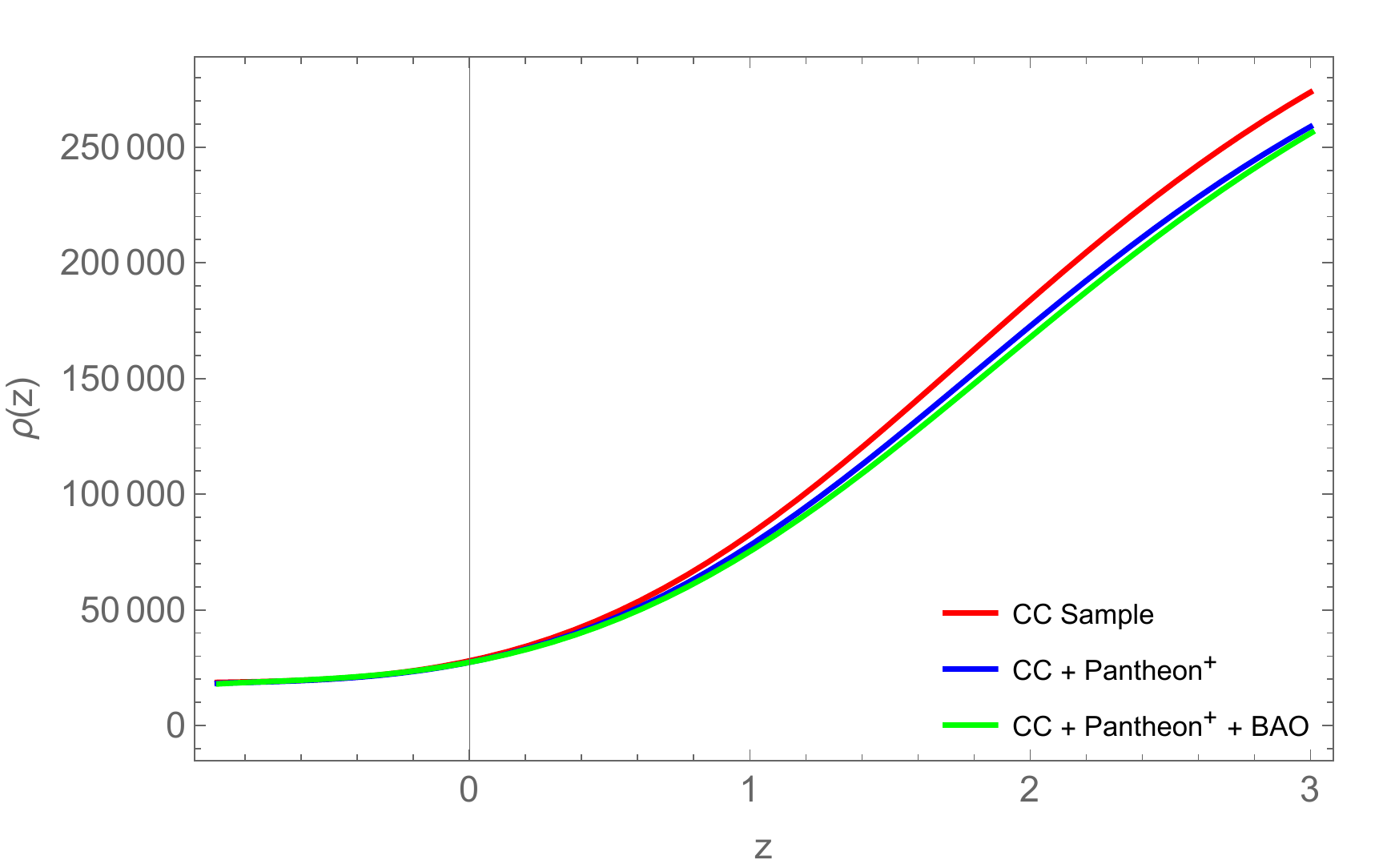}
\caption{Evolution of energy density in redshift. The parameter scheme: $\alpha_{2}=0.03$, $m=0.98$ and $\lambda=\kappa^{2}=1$.}
\label{fig:VII}
\end{figure}
\begin{figure}
\centering
\includegraphics[scale=0.5]{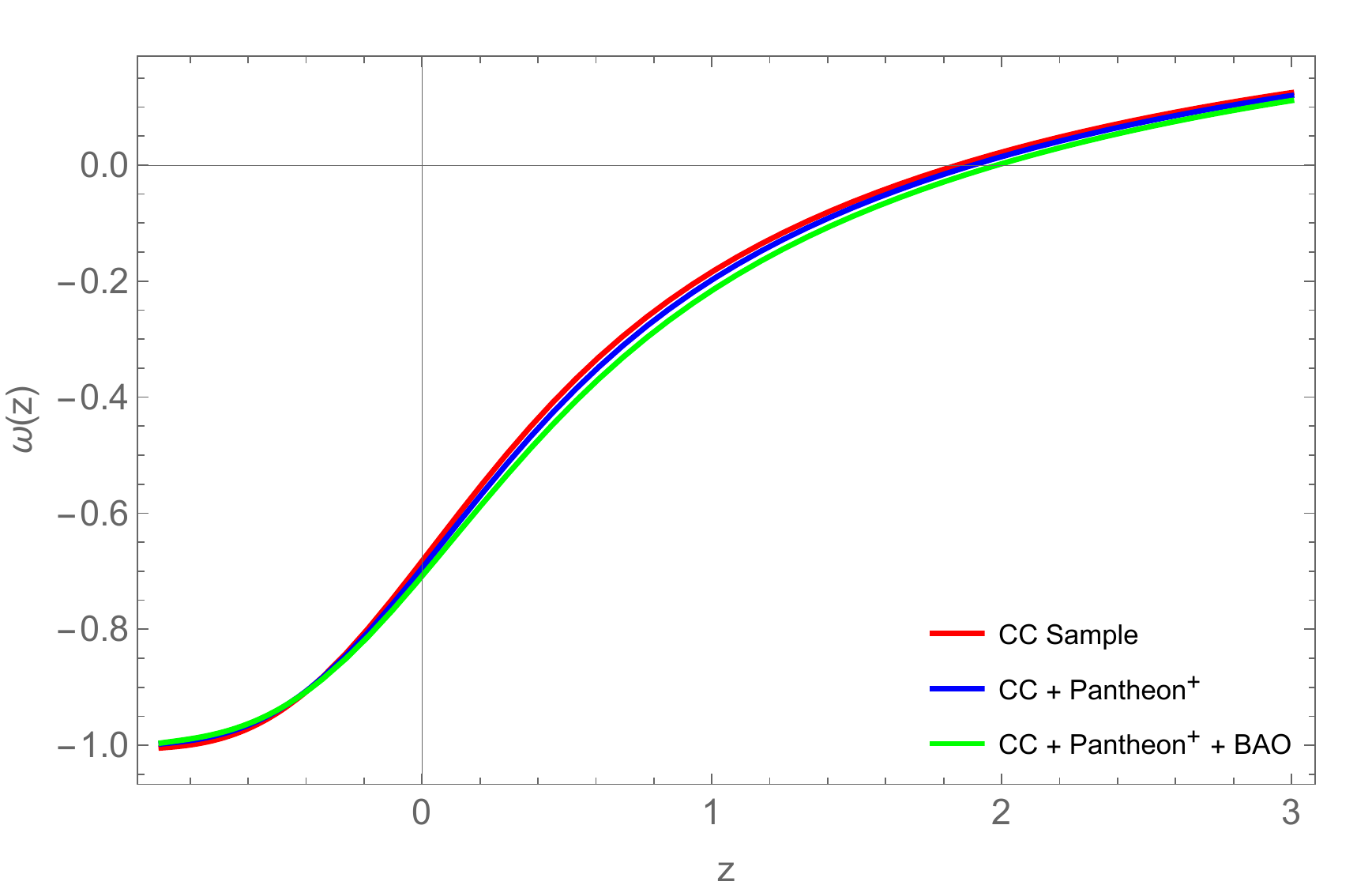}
\caption{Evolution of EoS parameter in redshift. The parameter scheme: $\alpha_{2}=0.03$, $m=0.98$ and $\lambda=\kappa^{2}=1$.}
\label{fig:VIII}
\end{figure}
\begin{figure}[H]
\centering
\includegraphics[scale=0.5]{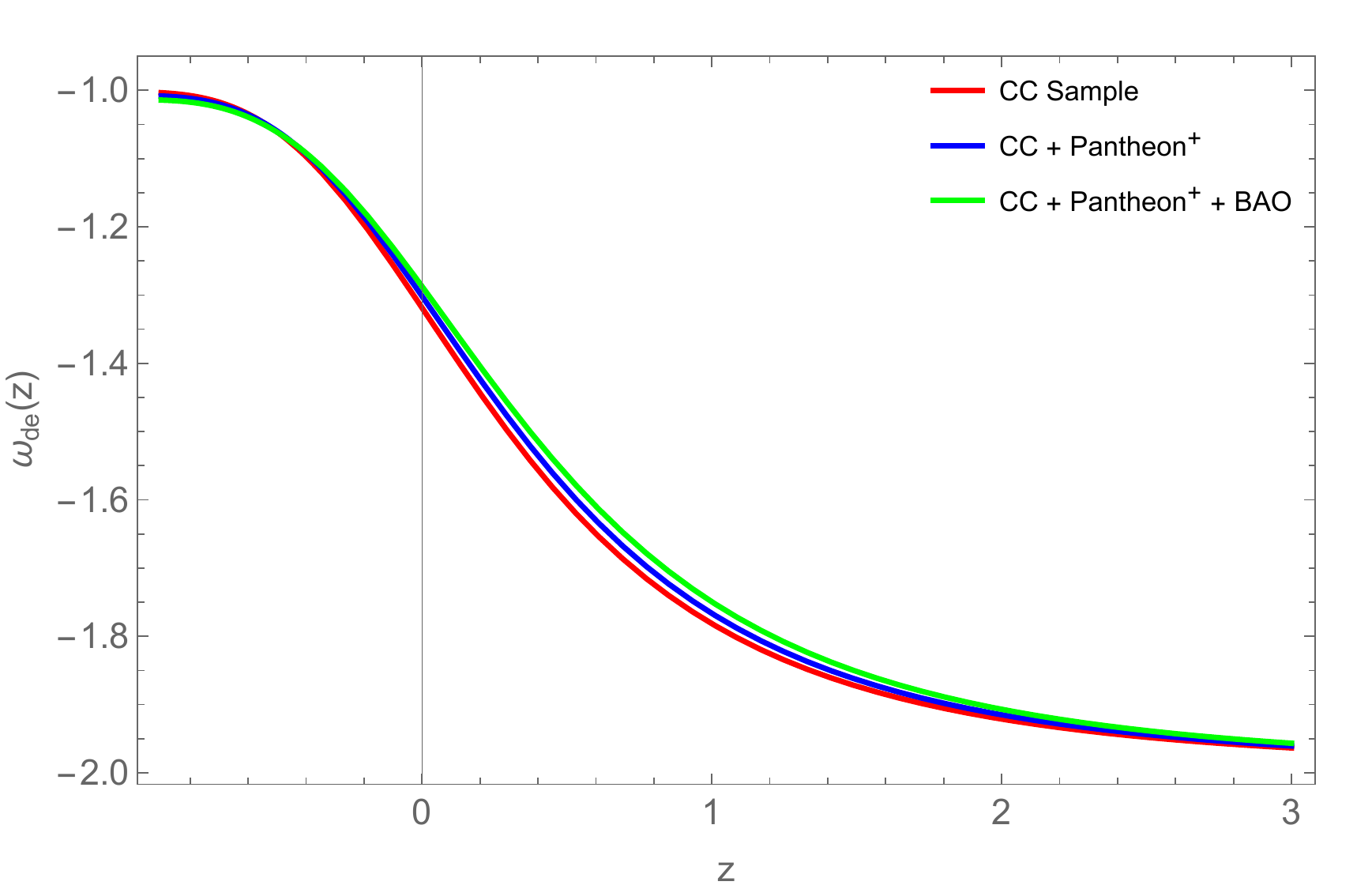}
\caption{Evolution of EoS parameter for dark energy in redshift. The parameter scheme: $\alpha_{2}=0.03$, $m=0.98$ and $\lambda=\kappa^{2}=1$.}
\label{fig:DEII}
\end{figure}

The value of free parameter $\alpha_{2}$ is chosen such that energy density shows decreasing behaviour and should be positive throughout the evolution [FIG.--\ref{fig:VII}]. The present value of EoS parameter obtained as, $\omega\approx -0.6828$, $\omega\approx -0.6949$ and $\omega\approx -0.7084$ respectively for $CC~Sample$, $CC + Pantheon^{+}$ and $CC + Pantheon^{+} + BAO$ datasets. The Universe shows quintessence behavior at present, while goes to $\Lambda$CDM in late time [FIG.--\ref{fig:VIII}]. The behaviour of EoS parameter for dark energy remains negative throughout the evolution. It shows increasing behaviour from early time to late time [see FIG. --\ref{fig:DEII}].

\section{Energy Conditions} \label{section 6}
Another important study of the cosmological model in modified theories of gravity is the behaviour of the energy conditions. We present here the behaviour of each of the energy conditions of both the models given in the Weyl type $f(Q,T)$ gravity. The general expressions for the energy conditions and their forms can be described as \cite{Novello2008},
\begin{itemize}
\item NEC (Null Energy Condition), for each null vector,
\begin{equation}
T_{ij}u^{i}u^{j} \geq 0 \Rightarrow \rho + p \geq 0 \Rightarrow  \geq 0.\label{eq:35}
\end{equation}
\item WEC (Weak Energy Condition), for every time-like vector,
\begin{equation}
T_{ij}u^{i}u^{j} \geq 0 \Rightarrow \rho \geq 0~~and~~ \rho + p \geq 0 .  \label{eq:36}
\end{equation}
\item SEC (Strong Energy Condition), for any time-like vector,
\begin{equation}
\left(T_{ij}-\frac{1}{2}Tg_{ij}\right)u^{i}u^{j} \geq 0 \Rightarrow  \rho + 3p \geq 0. \label{eq:37}
\end{equation}
\item DEC (Dominant Energy Condition), for any time like vector, 
\begin{equation}
T_{ij}u^{i}u^{j} \geq 0 \Rightarrow \rho - p \geq 0~~and~~ T_{ij}u^{j}~\text{not~spacelike}. \label{eq:38}
\end{equation}
\end{itemize}
 The energy density requires to be positive throughout the evolution and the energy conditions are effectively the boundary conditions. The violation of SEC due to the effect of DE, indicates that there is no physical reality to these boundary conditions. However, because of the fundamental casual structure of space time, the gravitation attraction can be characterized by the energy conditions \cite{Capozziello2019}. These boundary conditions are also useful in shaping the cosmic evolution of the Universe \cite{Carroll2003}. 
 
\begin{widetext}

\begin{figure}
\centering
\includegraphics[scale=0.5]{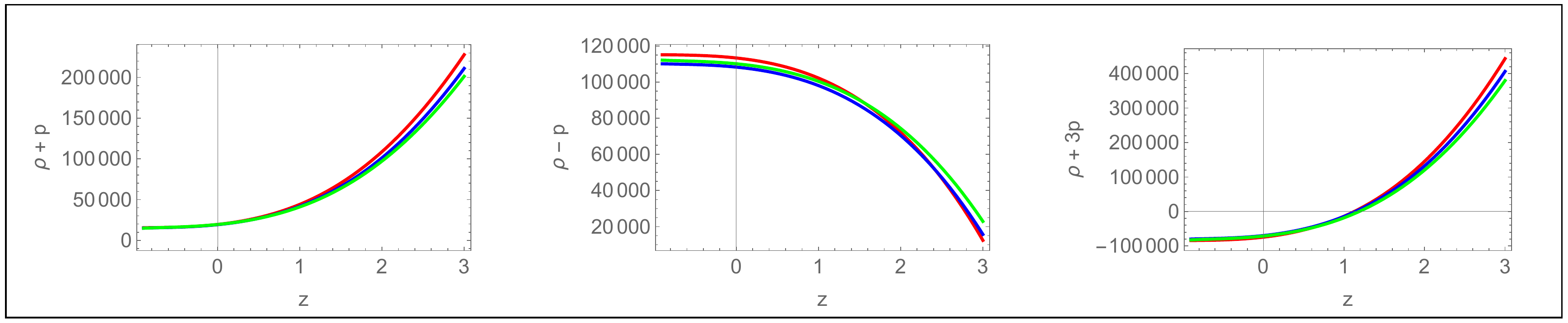}
\includegraphics[scale=0.5]{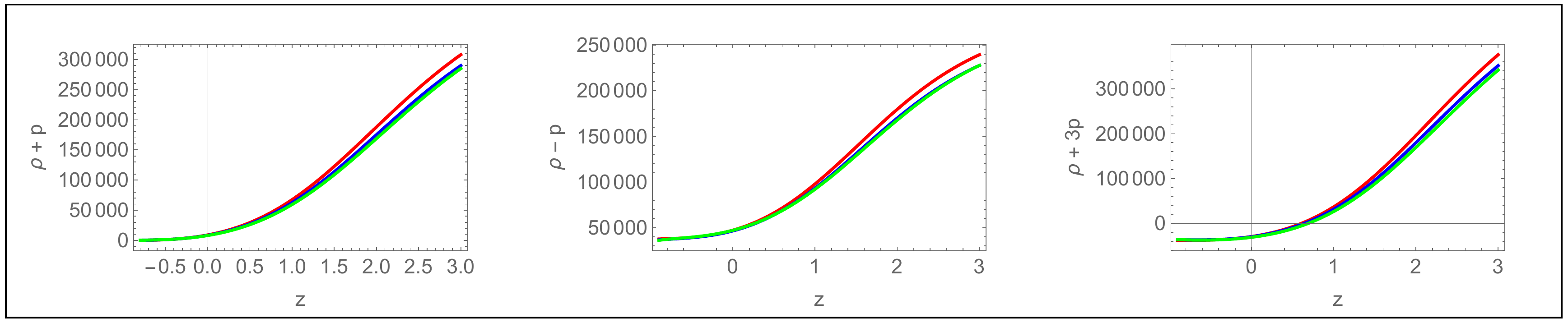}
\caption{Energy conditions in redshift. Model--I(Upper Panel) and Model--II (Lower Panel).}
    \label{fig:IX}
\end{figure}
\end{widetext}

We have given the expanded expressions of the energy conditions for both the models in the {\bf Appendix}. The behaviour of energy conditions for both models have been given in FIG.- \ref{fig:IX}. It has been observed the similar evolutionary behaviour of energy conditions in both the models. The NEC is decreasing from early epoch to late epoch, remains positive throughout, however at late time it vanishes. Whereas, DEC remains positive throughout and does not violate. The SEC satisfies at early time and violets at late time. 

\section{Conclusion}\label{section 7}
We have investigated the cosmological aspects of Weyl type $f(Q,T)$ gravity in an isotropic and homogeneous space time. The non-metricity scalar $Q$ has been expressed in its standard Weyl form and can be determined by the vector field $w_{\mu}$. From the well motivated log model parametrization expression, We have obtained the parametrized form of $H(z)$ and the free parameters are constrained using the cosmological datasets such as $32-$ points of $CC$ sample, $1701-$ light curves from $Pantheon^{+}$ dataset and $6-~BAO$ distance dataset. To note here this parametrization is a bit less 
flexible parameterization than that of the epsilon model \cite{Lemos2018}. However, the free parameters as given in Eqn. \eqref{eq:20} are less degenerate. We observed that the data constrain $H(z)$ to be so close to the form of  $\Lambda$CDM cosmology. Also for each dataset, we have analysed the behaviour of the deceleration parameter, which gives accelerating behaviour of the Universe and the state finder diagnostic pair $(j,~s)$, which confirms the  quintessence behaviour of the Universe. All the constrained values of free parameters are given in TABLE--\ref{table:I} and the present value of the cosmographic parameters along with the transition point are listed in TABLE-\ref{table:II}. \\

We have employed two well motivated forms of the function $f(Q,T)$ as (i) Exponential form leads to a complex cosmological dynamics, involving larger deviations from the $\Lambda$CDM model and; (ii) Non-linear form which gives us a good description of the standard cosmological model at small redshifts.  An energy density that is physically acceptable is achieved by choosing the model parameters appropriately, so that the positive value can be maintained throughout the evolution. In both the models the energy density  shows the decreasing behaviour over time.  The EoS parameter for the exponential model and non-linear model obtained as $\omega\approx -0.7068$ and $\omega\approx -0.6828$ for $CC~Sample$,  $\omega\approx -0.6991$ and $-0.6949$ for $CC + Pantheon^{+}$ and $\omega\approx -0.7001$ and $-0.7084$ for $CC + Pantheon^{+} + BAO$ respectively.  Apart from the present value of EoS from the cosmological datasets, in both the models the EoS parameter shows the transient behaviour i.e. an early time positive value to late time negative value. However, in the dark energy phase, it remains entirely in the negative region. We have analysed the energy conditions for both the models and as expected. the violation of SEC at late time for both models have been obtained. However the NEC and DEC does not violate, but NEC vanishes at late time.

Finally, we conclude that the parameterized $H(z)$ is consistent with current observational data and it can reproduce the $\Lambda$CDM cosmology. The $q(z)$ shows acceleration behavior of Universe with transition. We can remark that both the models support the quintessence behaviour at the present time which is verified by the state finder pair. Our work demonstrates that the Weyl type $f(Q,T)$ model can be useful in addressing the recent cosmic phenomena of the Universe.\\

\section*{Acknowledgement} RB acknowledges the financial support provided by University Grants Commission (UGC) through Junior Research Fellowship UGC-Ref. No.: 211610028858 to carry out the research work. BM acknowledges the support of IUCAA, Pune (India) through the visiting associateship program. The authors are thankful to the anonymous referees for their valuable comments and suggestions to improve the quality of the paper.

\begin{center}
\begin{table}[H]
\caption{The observational datasets that was used in \cite{Moresco2022}.} 
\centering 
\begin{tabular}{c c c c c | c c c c c} 
\hline\hline 
No. & Redshift & $H(z)$ & $\sigma_{H}$ & Ref. & No. & Redshift & $H(z)$ & $\sigma_{H}$ & Ref.\\ [0.5ex] 
\hline 
1.  & 0.07 & 69.0 & 19.6 & \cite{Zhang2014}      & 17. & 0.4783 & 80.9 & 9.0 &  \cite{Moresco2016} \\ 
2.  & 0.09 & 69.0 & 12.0 & \cite{Simon2005}    & 18. & 0.48 & 97 & 62 &  \cite{Stern2010} \\ 
3.  & 0.12 & 68.6 & 26.2 & \cite{Zhang2014}      & 19. & 0.593 & 104 & 13 & \cite{Moresco2012} \\
4.  & 0.17 & 83 & 8 &  \cite{Simon2005}          & 20. & 0.68 & 92 & 8 & \cite{Moresco2012} \\
5.  & 0.179 & 75.0 & 4.0 & \cite{Moresco2012}    & 21. & 0.75 & 98.8 & 33.6 &  \cite{Borghi2022} \\
6.  & 0.199 & 75.0 & 5.0 &  \cite{Moresco2012}   & 22. & 0.781 & 105 & 12 &  \cite{Moresco2012}\\
7.  & 0.200 & 72.9 & 29.6 &  \cite{Zhang2014}    & 23. & 0.875 & 125 & 17 &  \cite{Moresco2012} \\
8.  & 0.27 & 77 & 14 &  \cite{Simon2005}         & 24. & 0.88 & 90 & 40 &  \cite{Stern2010}  \\
9.  & 0.28 & 88.8 & 36.6 & \cite{Zhang2014}      & 25. & 0.9 & 117 & 23 &  \cite{Simon2005} \\
10. & 0.352 & 83 & 14 &  \cite{Moresco2012}      & 26. & 1.037 & 154 & 20 &  \cite{Moresco2012} \\ 
11. & 0.38 & 83.0 & 13.5 &  \cite{Moresco2016}   & 27. & 1.3 & 168 & 17 &  \cite{Simon2005} \\
12. & 0.4 & 95 & 17 &  \cite{Simon2005}          & 28. & 1.363 & 160 & 33.6 & \cite{Moresco2015} \\
13. & 0.4004 & 77 & 10.2 &  \cite{Moresco2016}   & 29. & 1.43 & 177 & 18 &  \cite{Simon2005} \\
14. & 0.425 & 87.1 & 11.2 &  \cite{Moresco2016}  & 30  & 1.53 & 140 & 14 &  \cite{Simon2005} \\
15. & 0.445 & 92.8 & 12.9 &  \cite{Moresco2016}  & 31. & 1.75 & 202 & 40 &  \cite{Simon2005} \\
16. & 0.47 & 89 & 49.6 & \cite{Ratsimbazafy2017} & 32. & 1.965 & 186.5 & 50.4 & \cite{Moresco2015} \\ 

\hline 
\end{tabular}
\label{table:III} 
\end{table}
\end{center}
\begin{widetext}

\section*{{\bf Appendix}}
 \textbf{Model I} \\
 \begin{eqnarray*}
     \rho+p &=& \frac{H^2\left(-38.8812\beta_{1}+\alpha_{1}(-6\beta_{1}-9)\mu e^{\frac{H^2 \mu}{H_{0}^2}}-58.3218\right)+(-12\beta_{1}-18)\dot{H}}{\beta_{1}(\beta_{1}+4.5)+4.5}~,\\
    \rho - p &=& \frac{(18\beta_{1}+54)H^2+\alpha_{1}(\beta_{1}+3) e^{\frac{H^2\mu}{H_{0}^2}} \left(3H_{0}^2-3H^{2}\mu\right)+(6\beta_{1}+18)\dot{H}}{\beta_{1}(\beta_{1}+4.5)+4.5}~,\\
    \rho+3p &=& \frac{(-95.7624 \beta_{1}-170.644) H^2+\alpha_{1}e^{\frac{H^{2}\mu}{H_{0}^2}} \left((-9\beta_{1}-9)H^2\mu +(-3\beta_{1}-9)H_{0}^2\right)+(-30\beta_{1}-54)\dot{H}}{\beta_{1}(\beta_{1}+4.5)+4.5}~.
\end{eqnarray*}

 \textbf{Model II} \\
 \begin{eqnarray*}
     \rho+p &=& \frac{H_{0}^2 \left(48\alpha_{2}H^4\lambda-H^2H_{0}^2 \left(-12 \kappa^2+12 \lambda +m^2\right)-4 \lambda\dot{H}\left(H_{0}^2-4\alpha_{2} H^2\right)\right)}{8 \alpha_{2}^{2}H^4+4\alpha_{2}H^2H_{0}^2+H_{0}^4}~,\\
     \rho - p &=& \frac{2H_{0}^2 \left(\alpha_{2}H^4\left(-12\kappa^2+36\lambda +m^2\right)+6H^{2}H_{0}^2\lambda +2\lambda\dot{H}\left(6\alpha_{2}H^2+H_{0}^2\right)\right)}{8\alpha_{2}^2 H^4+4\alpha_{2}H^2H_{0}^2+H_{0}^4}~,\\
     \rho+3p &=& -\frac{2H_{0}^2\left(\alpha_{2}H^4\left(m^2-12\left(\kappa ^2+\lambda\right)\right)+H^2H_{0}^2\left(-12\kappa ^2+18\lambda +m^2\right)+\dot{H}\left(6H_{0}^2\lambda -4 \alpha_{2}H^2\lambda \right)\right)}{8 \alpha_{2}^2 H^4+4\alpha_{2}H^2H_{0}^2+H_{0}^4}~.
 \end{eqnarray*}
\end{widetext}

\end{document}